\documentclass[preprint,amssymb,nofootinbib,nobibnotes,showpacs,groupaddress,preprintnumbers,amsmath,aps,prd]{revtex4}
\usepackage{epsfig,amsmath,graphicx,amssymb}
\usepackage{graphicx}
\usepackage{dcolumn}
\usepackage{bm}
\begin{document}
\renewcommand{\pl}{\partial}
\newcommand{\be}[0]{\begin{equation}}
\newcommand{\ee}[0]{\end{equation}}
\newcommand{\f}[2]{\frac{#1}{#2}}
\def\bea{\begin{eqnarray}}
\def\eea{\end{eqnarray}}
\def\bes{\begin{subequations}}
\def\ees{\end{subequations}}

\title{Guiding ultraslow weak-light bullets with Airy beams
in a coherent atomic system}
\author{Chao Hang}
\email[]{chang@phy.ecnu.edu.cn}
\affiliation{State Key Laboratory of Precision Spectroscopy and
Department of Physics, East China Normal University, Shanghai
200062, China}
\author{Guoxiang Huang}
\email[]{gxhuang@phy.ecnu.edu.cn}
\affiliation{State Key Laboratory of Precision Spectroscopy and
Department of Physics, East China Normal University, Shanghai
200062, China}
\date{\today}

\begin{abstract}

We investigate the possibility of guiding stable ultraslow
weak-light bullets by using Airy beams in a cold,
lifetime-broadened four-level atomic system via
electromagnetically induced transparency (EIT). We show that under
EIT condition the light bullet with ultraslow propagating velocity
and extremely low generation power formed by the balance between
diffraction and nonlinearity in the probe field can be not only
stabilized but also steered by the assisted field. In particular,
when the assisted field is taken to be an Airy beam, the light
bullet can be trapped into the main lobe of the Airy beam,
propagate ultraslowly in longitudinal direction, accelerate in
transverse directions, and move along a parabolic trajectory. We
further show that the light bullet can bypass an obstacle when
guided by two sequential Airy beams. A technique for generating
ultraslow helical weak-light bullets is also proposed.

\end{abstract}


\maketitle

\section{Introduction}

In the past two decades, much effort has been paid to
study of spatial-temporal optical solitons, or light bullets,
which describe a fascinating class of nonlinear optical pulses
localized in three spatial and one temporal dimensions~\cite{sil}.
Due to the balance between diffraction, dispersion,
and nonlinearity, these optical pulses are capable of arresting
spatial-temporal distortion and propagate stably for a long
distance. Light bullets are of great interest because of their
rich nonlinear physics and important applications
\cite{ber,kiv0,mal,kiv,liu,bla,tow,tra,mih,mat,ber1,bel,burg,chen,abd,min,mat1,kat1,mihalache1,mihalache2,mihalache3}.
However, up to now most light bullets are produced in passive
optical media, in which far-off resonance excitation schemes are
employed in order to avoid significant optical absorption. For
generating the light bullets in passive media, very high
light-intensity is usually needed to obtain nonlinearity strong
enough to balance the dispersion and diffraction effects. In
addition, an active control on the property of light bullets is
not easy to realize in passive media because of the absence of
energy-level structure and selection rules that can be used and
manipulated.

For practical applications, light bullets having low
generation power and good controllability are highly desirable.
Active optical media, in which light interacts with matter
resonantly, can be adopted to achieve such goal. However, in
resonant media there is usually a large optical
absorption. In order to suppress the large optical absorption,
a technique called electromagnetically induced transparency (EIT)
\cite{har} can be used. Due to the quantum interference effect induced by a
control field, the propagation of a weak probe field
in EIT media exhibits not only large suppression of optical absorption,
but also significant
reduction of group velocity, and great enhancement of
Kerr nonlinearity, etc~\cite{fle}. Based on these important features,
new types of temporal \cite{wu,hua,hang,yang} and spatial
\cite{hong,mic,hang1,hang2} optical solitons were predicted
in highly resonant atomic systems via EIT. The existence of
ultraslow light bullets was also demonstrated~\cite{LWH}.
Active control of these optical solitons by using
Stern-Gerlach gradient magnetic fields were also
explored recently~\cite{han1,han2}.

In this article, we investigate how to guide stable
ultraslow weak-light bullets by means of Airy beams in a cold,
lifetime-broadened four-level atomic system via EIT. Under EIT
condition, assisted-field envelope obeys a (2+1)-dimensional
linear Helmholtz equation supporting Airy beam solutions, which
contributes a trapping potential to probe-field envelope governed
by a (3+1)-dimensional nonlinear Schr\"{o}dinger equation. We show
that, both analytically and numerically, the light bullet with
ultraslow propagating velocity ($\sim 10^{-5}\,c$; $c$ is the
light speed in vacuum) and extremely low generation power ($\sim
1\,\mu W$) formed by the balance between diffraction and
nonlinearity in the probe field can be not only stabilized but
also guided by the assisted field. In particular, when the
assisted field is taken to be an Airy beam the light bullet can be
trapped into the main lobe of the Airy beam, propagate ultraslowly
in longitudinal direction, accelerate in transverse directions, and hence move
along a parabolic trajectory. Interestingly, the light
bullet can bypass an obstacle when guided by two sequential Airy
beams. In addition, a technique of generating ultraslow helical weak-light
bullets using sequential Airy and Bessel beams is proposed. The results
presented here are useful for guiding new experimental
findings and have potential applications in optical information
processing and transmission.

Before proceeding, we note that due to the pioneering work by
Berry and Balazs~\cite{Berry}, recently there is growing interest
focused on the study of Airy beams. Due to their unique
interference, Airy beams undergo no temporal spreading (spatial
diffraction) and have the ability to freely accelerate (bend)
requiring no waveguiding structures or external
potentials~\cite{Bandres}. In addition to fundamental research
interest, accelerating Airy beams have led to many intriguing
ideas and exciting applications, including particle and cell
micromanipulation, laser micromachining, generation of curved
plasma channel, generation of curved electron beams, and so
on~\cite{bau,zhang,Polynkin,zhu}. Different from the previous
studies, where Airy light beams have been used to manipulate the
movement of material (or massive) particles, in our work the
particles are not material ones but light wavepackets (light
bullets), which are steered by using Airy light beams in a highly
controllable way. To the best of our knowledge, no such study has
been reported up to now.

The article is arranged as follows. In the next section, we
introduce the model and deduce the nonlinear envelope equations
governing the envelopes of probe and assisted fields. In Sec.~III, we
investigate the guiding of ultraslow weak-light bullets with Airy
beams. We also demonstrate that the light bullet can bypass
an obstacle when it is guided by two sequential Airy beams. In
Sec.~IV, generation of ultraslow helical weak-light bullets is
discussed. Finally, in the last section we summarize the main results
obtained in this work.

\section{Model and nonlinear envelope equations}
\label{sec2}

\subsection{Model}

\begin{figure}
\centering
\includegraphics[scale=0.5]{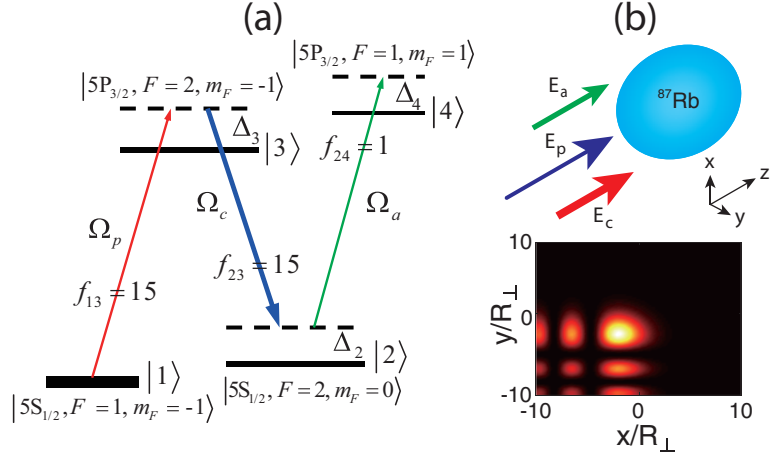}
\caption{\footnotesize (Color online) (a) Energy-level diagram
and excitation scheme of the lifetime-broadened four-state atomic
system interacting with a weak pulsed probe field (with half Rabi
frequency $\Omega_{p}$),  a strong CW control field (with half
Rabi frequency $\Omega_{c}$), and a weak CW assisted field (with
half Rabi frequency $\Omega_{a}$). $\Delta_3,$ $\Delta_{2}$, and
$\Delta_{4}$ are one-photon, two-photon, and three-photon
detunings, respectively.  The energy levels are taken from the
D$_{2}$ line of $^{87}$Rb atoms, with $|1\rangle=|5 {\rm S}_{1/2},
F=1, m_F=-1\rangle$, $|2\rangle=|5 {\rm S}_{1/2}, F=2,
m_F=0\rangle$, $|3\rangle=|5 {\rm P}_{3/2}, F=2, m_F=-1\rangle$,
and $|4\rangle=|5 {\rm P}_{3/2}, F=1, m_F=1\rangle$.
$f_{ij}=|\textbf{p}_{ij}/D2|^{2}\times120$ is the relative
transition strength, with $D2=3.58\times10^{-27}$ cm C and
$\textbf{p}_{ij}$ being the dipole transition matrix element
between the state $|i\rangle$ and the state $|j\rangle$. (b)
The geometry of the system. The lower part shows the intensity pattern of
the assisted field, chosen as an Airy beam, in the
$x$-$y$ plane. } \label{fig1}
\end{figure}

We consider a cold, lifetime broadened atomic system with N-type
energy-level configuration, shown in Fig.~\ref{fig1}(a).
%
%
A weak, pulsed probe field (strong, continuous-wave (CW) control field) with angular
frequency $\omega_{p}$ ($\omega_{c}$) and wavevector
$\textbf{k}_{p}$\, ($\textbf{k}_{c}$) interacts resonantly with
the energy states $|1\rangle$ and $|3\rangle$ ($|2\rangle$ and
$|3\rangle$). In addition, a weak assisted laser field
with angular frequency $\omega_{a}$ and wavevector
$\textbf{k}_{a}$ couples to energy states $|2\rangle$ and
$|4\rangle$, which contributes a cross-phase modulation (CPM) to
the probe field, as shown below. The energy levels can be selected from the
D$_{2}$ line of $^{87}$Rb atoms, with the states assigned as
$|1\rangle=|5 {\rm S}_{1/2}, F=1, m_F=-1\rangle$, $|2\rangle=|5
{\rm S}_{1/2}, F=2, m_F=0\rangle$, $|3\rangle=|5 {\rm P}_{3/2},
F=2, m_F=-1\rangle$, and $|4\rangle=|5 {\rm P}_{3/2}, F=2,
m_F=1\rangle$ (see Fig. \ref{fig1}). In the figure, $f_{ij}$ is the relative
transition strength, defined by
$f_{ij}=|\textbf{p}_{ij}/D2|^{2}\times120$. Here
$D2=3.58\times10^{-27}$ cm C and $\textbf{p}_{ij}$ is the dipole
transition matrix element between the state $|i\rangle$, and the
state $|j\rangle$ \cite{steck}. The electric-field vector in the
system can be written as
$\textbf{E}=\sum_{l=p,c,a}\textbf{e}_{l}{\cal
E}_{l}\exp{[i(\textbf{k}_{l} \cdot\textbf{r}-\omega_{l}t)]}+{\rm
c.c.}, $  where $\textbf{e}_{l}$ is polarization direction of
$l$th field with envelope ${\cal E}_{l}$. The geometry of the
system is illustrated in Fig.~\ref{fig1}(b).

Under electric-dipole and rotating-wave approximations, the
Hamiltonian in the interaction picture reads
$ \hat{H}_{\rm int}=-\hbar\sum_{j=1}^{4}\Delta_{j}|j\rangle\langle
j|-\hbar(\Omega_{p}|3\rangle\langle 1|+\Omega_{c}|3\rangle\langle
2|+\Omega_{a}|4\rangle\langle 2|+{\rm H. c.}),$
where $\Delta_{3}=\omega_{p}-(\omega_{3}-\omega_{1}),$
$\Delta_{2}=\omega_{p}-\omega_{c}-(\omega_{2}-\omega_{1}),$ and
$\Delta_{4}=\omega_{p}-\omega_{c}+\omega_{a}-(\omega_{4}-\omega_{1})$
are respectively the one-, two-, and three-photon detunings.
$\Omega_{p}=(\textbf{e}_{p}\cdot\textbf{p}_{13}){\cal
E}_{p}/\hbar$,
$\Omega_{c}=(\textbf{e}_{c}\cdot\textbf{p}_{23}){\cal
E}_{c}/\hbar$, and $\Omega_{a}=
(\textbf{e}_{a}\cdot\textbf{p}_{24}){\cal E}_{a}/\hbar$ are respectively half
Rabi frequencies of the probe, control, and assisted fields.

The equation of motion for the density-matrix $\sigma$ reads
\begin{equation}
\frac{\pl \sigma}{\pl t}=-\frac{i}{\hbar}\left[\hat{H}_{\rm int},\sigma\right]-\Gamma \sigma,
\end{equation}
where $\Gamma$ is a $4\times 4$ relaxation matrix. Explicit expressions of the
equations of motion for $\sigma_{ij}$ have been given in the Appendix A.

Electric-field evolution is controlled by Maxwell equation
$\nabla^{2}\textbf{E}-(1/c^{2})\pl^{2}\textbf{E}/\pl t^{2}
=(1/\epsilon_{0}c^{2})\pl^{2}\textbf{P}/\pl t^{2}$,
with
$
\textbf{P}=N\{\textbf{p}_{13}\sigma_{31}\exp[i(\textbf{k}_{p}\cdot
\textbf{r}-\omega_{p}t)]
+\textbf{p}_{23}\sigma_{32}\exp[i(\textbf{k}_{c}\cdot
\textbf{r}-\omega_{c}t)]+\textbf{p}_{24}\sigma_{42}
\exp[i(\textbf{k}_{a}\cdot \textbf{r}-\omega_{a}t)]+{\rm c.c.}\}$.
Under a slowly varying envelope approximation, we obtain the
equations for $\Omega_{p}$ and $\Omega_{a}$:
\bes\label{max}
\bea
& &i\left(\f{\pl}{\pl z}+\f{1}{c}\frac{\pl}{\pl
t}\right)\Omega_{p}+\f{c}{2\omega_{p}}\left(\f{\pl ^{2}}{\pl
x^{2}}+\f{\pl ^{2}}{\pl y^{2}}\right)\Omega_{p}+\kappa_{13} \sigma_{31}=0,\label{max1}\\
& &i\f{\pl}{\pl z}\Omega_{a}+\f{c}{2\omega_{a}}\left(\f{\pl
^{2}}{\pl x^{2}}+\f{\pl ^{2}}{\pl
y^{2}}\right)\Omega_{a}+\kappa_{24}\sigma_{42}=0,\label{max2}
\eea
\ees
where
$\kappa_{13,24}=N\omega_{p,a}|\textbf{e}_{p,a}\cdot\textbf{p}_{13,24}|^2/(2\epsilon_{0}\hbar
c)$, with $N$ being atomic concentration. For simplicity, the
probe field and the assisted field have been assumed to propagate
in $z$-direction, i.e. $\textbf{k}_{p,a}=\textbf{e}_{z}k_{p,a}$.

\subsection{Asymptotic expansion and nonlinear envelope equations}

Because we are interested in the nonlinear evolution and the
possible formation of optical solitons in the system, we employ
the standard method of multiple-scales, to investigate the
evolution of both the probe and assisted fields. The atoms are
assumed to be initially populated in the state $|1\rangle$. We
make the asymptotic expansions
$\sigma_{ij}=\sigma_{ij}^{(0)}+\epsilon \sigma_{ij}^{(1)}+
\epsilon^{2}\sigma_{ij}^{(2)}+\epsilon^{3}\sigma_{ij}^{(3)}
+\cdots$,  and $\Omega_{p,a}=\epsilon \Omega_{p,a}^{(1)}
+\epsilon^{2}\Omega_{p,a}^{(2)}+\epsilon^{3}\Omega_{p,a}^{(3)}
+\cdots$, with $\sigma_{ij}^{(0)}=\delta_{i1}\delta_{j1}$ (both
$\delta_{i1}$ and $\delta_{j1}$ are Kronecker delta symbols). Here
$\epsilon$ is a small parameter characterizing the typical
amplitude of the probe and assisted fields. To obtain
divergence-free expansions, all quantities on the right hand sides
of the asymptotic expansions are considered as functions of the
multi-scale variables $z_{l}=\epsilon^{l}z$ ($l=0, 1, 2$),
$t_{l}=\epsilon^{l}t$ ($l=0, 2$), $x_{1}=\epsilon x$, and
$y_{1}=\epsilon y$. Substituting these expansions into Eqs.
(\ref{den}) and (\ref{max}), one can obtain a series of linear but
inhomogeneous equations for $\sigma_{ij}^{(l)}$ and
$\Omega_{p,a}^{(l)}$ ($l=1,2,3,...$), which can be solved order by
order.

At the first-order, we obtain the solution under linear level:
\bes\label{jeq1}
\bea
& & \Omega_{p}^{(1)}=F\,e^{i\theta},\,\,\Omega_{a}^{(1)}=G,\\
& &
\sigma_{j1}^{(1)}=\f{-\delta_{j2}\Omega_{c}^{\ast}+\delta_{j3}(\omega+d_{21})}{D}F
e^{i\theta},\,\,(j=2,3)
\eea
\ees
with $D=|\Omega_{c}|^{2}-(\omega+d_{21})(\omega+d_{31})$, and other $\sigma_{ij}^{(1)}$ being zero. In the
above expressions, $\theta=K(\omega)z_{0}-\omega t_{0}$, $F$ and
$G$ are yet to be determined envelope functions depending on the
slowly-varying variables $t_2$, $z_1$, and $z_2$. We see that in this order
the two weak fields evolve independently. Moreover, the assisted
field is free, but the probe field experiences a dispersion and
absorption obeying the linear dispersion relation:
\be\label{disp}
K(\omega)=\f{\omega}{c}+\kappa_{13}\f{\omega+d_{21}}{D}. \ee
%

%
%
\begin{figure}
\centering
\includegraphics[scale=0.6]{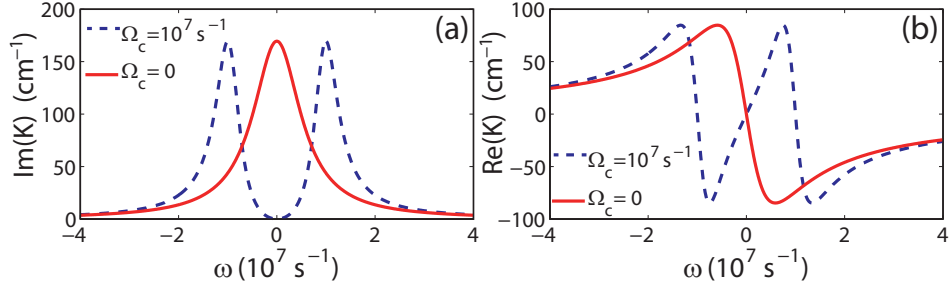}
\caption{\footnotesize (Color online) The imaginary part
Im$K(\omega)$ (panel (a)\,) and the real part Re$K(\omega)$ (panel
(b)\,) of the linear dispersion relation $K(\omega)$ of the probe field as
functions of  $\omega$. In both panels, the
dashed and solid lines correspond to the presence
($\Omega_c=1.0\times 10^7$ s$^{-1}$) and the absence
($\Omega_c=0$) of the control field, respectively. The other
parameters are given in the text. A transparency window is opened
for large control field (the dashed line in panel (a)\,). The
steep slope of the dashed-dotted line for large control field (the
dashed line in panel (b)\,) results in a ultraslow group velocity.}
\label{fig2}
\end{figure}
%
Shown in Fig.~\ref{fig2} is the imaginary part Im$K(\omega)$
(Fig.~\ref{fig2}(a)\,) and the real part Re$K(\omega)$ (Fig.~\ref{fig2}(b)\,) of
$K(\omega)$ as functions of frequency $\omega$. As an example, we
take the system parameters as $\Gamma_{12}\approx1 \,{\rm
kHz},\,\Gamma_{13}\approx\Gamma_{23}\approx6\,{\rm
MHz},\,\Delta_{2,3}=0$ s$^{-1},$ and $\kappa_{13}=1.0\times 10^{9}
\, {\rm cm}^{-1} {\rm s}^{-1}.$ The dashed and the solid lines in
both panels correspond to the presence ($\Omega_c=1.0\times 10^7$
s$^{-1}$) and the absence ($\Omega_c=0$)\,) of the control field,
respectively. One sees that when $\Omega_c$ is absent, the probe
field has a large absorption (the solid line of Fig.~\ref{fig2}(a)\,);
however, when $\Omega_c$ is applied an EIT  transparency window is
opened (the dashed line of Fig.~\ref{fig2}(a)\,). The steep slope for large
control field (the dashed line of Fig.~\ref{fig2}(b)\,) results in a slow
group velocity at the center frequency of the probe field (i.e.
$\omega=0$ \cite{note1}). The suppression of the absorption and the reduction
of the group velocity are due to the EIT effect
induced by the control field.

At the second-order, the solvability condition for
$\sigma_{ij}^{(2)}$ and $\Omega_{p,a}^{(2)}$ requires $\pl F/\pl
z_{1}=0$ and $\pl G/\pl z_{1}=0$, and hence both $F$ and $G$ are
independent of $z_1$. At the third-order, using the solvability
condition for $\sigma_{ij}^{(3)}$ and $\Omega_{p,a}^{(3)}$ we
obtain the coupled nonlinear equations for $F$ and $G$:
\bes\label{solcon2}
\bea
& & \label{solcon21} i\left(\f{\pl }{\pl
z_2}+\f{1}{V_{g}}\f{\pl}{\pl
t_2}\right)F+\f{c}{2\omega_{p}}\left(\f{\pl ^{2}}{\pl
x_1^{2}}+\f{\pl ^{2}}{\pl y_1^{2}}\right)F
+\alpha_{11}|F|^{2}F+\alpha_{12}|G|^{2}F=0,\\
& & \label{solcon22} i\f{\pl}{\pl
z_{2}}G+\f{c}{2\omega_{a}}\left(\f{\pl ^{2}}{\pl x_1^{2}}+\f{\pl
^{2}}{\pl y_1^{2}}\right)G+\alpha_{21}|F|^{2}G=0,
\eea
\ees
where $V_g= ({\pl K}/{\pl \omega})^{-1}$ is the group velocity of
the envelope $F$. The explicit expressions for the coefficient of
self-phase modulation (SPM) of the probe field (i.e.
$\alpha_{11}$), and the CPM coefficients between the two fields (i.e. $\alpha_{12}$ and
$\alpha_{21}$), have been given in the Appendix B.

Since the selected atomic transition between $|2\rangle$ and $|4\rangle$
is much weak than those between $|1\rangle$ and
$|3\rangle$ and between $|2\rangle$ and $|3\rangle$, the
coupling constants in Eq.~(\ref{max}) satisfy $\kappa_{24}\ll
\kappa_{13}$, and hence $\alpha_{21}\ll \alpha_{11}, \alpha_{12}$.
In this way the CPM term in Eq.~(\ref{solcon22}) can be safely
neglected. Under this condition, Eq.~(\ref{solcon22}) is reduced
into a linear Helmholtz equation. As a result, we obtain the following
envelope equations
\bes\label{GNLSE1}
\bea
& & \label{GNLSE11}i\left(\f{\pl }{\pl z}+\f{1}{V_{g}}\f{\pl}{\pl
t}\right)U+\f{c}{2\omega_{p}}\left(\f{\pl ^{2}}{\pl x^{2}}+\f{\pl
^{2}}{\pl y^{2}}\right)U+\alpha_{11}|U|^{2}U+\alpha_{12}|V|^{2}U=0,\\
& & \label{GNLSE12} i\f{\pl V}{\pl z}+\f{c}{2\omega_{a}}\left(\f{\pl ^{2}}{\pl
x^{2}}+\f{\pl ^{2}}{\pl y^{2}}\right)V=0,
\eea
\ees
after returning to the original variables, where $U=\epsilon F$
and $V=\epsilon G$. One sees that the role of the assisted-field
envelope $V$ is now acting as an external potential (controlled by
Eq.~(\ref{GNLSE12})\,) to the probe field envelope $U$ (controlled
by Eq.~(\ref{GNLSE11})\,). This is desirable because the external
potential $|V|^2$ can be used not only to stabilize the motion of
$U$ but also to guide it along a particular path, as shown below.


\section{Guiding ultraslow weak-light bullets with Airy beams}

\subsection{Estimation on the coefficients in the nonlinear
envelope equations}

Before solving Eqs.~(\ref{GNLSE11}) and (\ref{GNLSE12}), we first
make an estimation on their coefficients by using realistic
physical parameters. Equations (\ref{GNLSE11}) and (\ref{GNLSE12})
can be written into the dimensionless form
\bes\label{GNLSE2}
\bea
& &\label{GNLSE21} i\left(\f{\pl }{\pl s}+\lambda\f{\pl }{\pl
\tau}\right)u+\f{1}{2}\left(\f{\pl ^{2}}{\pl \xi^{2}}+\f{\pl
^{2}}{\pl
\eta^{2}}\right)u+g_{11}|u|^{2}u+g_{12}v_{0}^2|v|^2u=0,\\
& &\label{GNLSE22} i\f{\pl v}{\pl s}+\f{\delta}{2}\left(\f{\pl ^{2}}{\pl \xi^{2}}+\f{\pl
^{2}}{\pl \eta^{2}}\right)v=0,
\eea
\ees
where $u=U/U_{0}$, $v=V/(U_{0}v_0)$, $s=z/L_{\rm diff}$,
$\lambda=L_{\rm Diff}/(V_g\tau_0)$, $\tau=t/\tau_0$ (with $\tau_0$
being the typical probe pulse length),
$(\xi,\eta)=(x,y)/R_{\perp}$ (with $R_{\perp}$ being the typical
probe beam radius), $g_{11}=\ell\alpha_{11}/|\alpha_{11}|$,
$g_{12}=\ell\alpha_{12}/|\alpha_{11}|$, and
$\delta=\omega_{p}/\omega_{a}$. Here $\ell=L_{\rm Diff}/L_{\rm
NL}$, with $L_{\rm Diff}\equiv \omega_{p}R_{\perp}^{2}/c$ being
the typical diffraction length, $L_{\rm
NL}=1/(|\alpha_{11}U_{0}^{2}|)$ being the typical nonlinear length,
and $U_0$ being the typical Rabi frequency of the probe field. The
typical Rabi frequency of the probe field can be solved as
$U_{0}=\sqrt{c/(\omega_{p}R_{\perp}^{2}|\alpha_{11}|)}$ if we take
$\ell=1$, i.e. take $L_{\rm Diff}=L_{\rm NL}$. $v_0$ is proportional to
the typical Rabi frequency of the assisted field, which is a free
parameter that can be used to adjust the magnitude of the
CPM coefficient, and hence control the stability of $U$.

Because the system we consider is lifetime-broadened, the
coefficients in the Eq.~(\ref{GNLSE21}) are generally complex. If
the control field Rabi frequency $\Omega_c$ is small, the
imaginary part of the coefficients is comparable with their real
part, and hence stable light bullet solutions do not exist.
However, under EIT condition $|\Omega_c|^2\gg
\gamma_{31}\gamma_{21}$ \cite{lil} the absorption of the probe
field can be largely suppressed, and hence the imaginary part of
these coefficients can be made to be much smaller than their real
part.

To show this we calculate the values of coefficients in the Eqs.
(\ref{GNLSE21}) and (\ref{GNLSE22}) by considering a cold atomic gas
of $^{87}$Rb atoms, with ${\rm D}_{2}$ line transitions $5^{2}{\rm
S}_{1/2}\rightarrow 5^{2}{\rm P}_{3/2}$. The energy levels are
chosen as those in Fig.~\ref{fig1}. From the data of $^{87}$Rb
\cite{steck}, we have the dipole matrix elements
$|\textbf{p}_{13}|\approx|\textbf{p}_{23}|=-\sqrt{\f{1}{8}}\times
3.58\times 10^{-27}$ cm C and
$|\textbf{p}_{24}|=\sqrt{\f{1}{120}}\times 3.58\times 10^{-27}$ cm
C. The other system parameters are taken as $\Gamma_{12}=1\,{\rm
kHz}$, $\Gamma_{13}\approx \Gamma_{23}\approx \Gamma_{24}/2=
35\,\,{\rm MHz}$, $\kappa_{13}=1.0\times 10^{10}\,\,{\rm cm}^{-1}
{\rm s}^{-1}$, $\kappa_{24}=1.0\times 10^{9}\,\, {\rm cm}^{-1}
{\rm s}^{-1}$, $\Omega_{c}=5.0\times10^{7}\,{\rm s}^{-1}$,
$\Delta_{2}=-1.5\times10^{6}$ s$^{-1}$,
$\Delta_{3}=-3.0\times10^{8}\,{\rm s}^{-1}$,
$\Delta_{4}=-1.0\times 10^{9}$ s$^{-1}$, $R_{\perp}=4.0\times
10^{-3}\,{\rm cm}$, and $U_{0}=9.0\times10^{6}$ s$^{-1}$. Then we
have $\delta\approx1.0$, $g_{11}\approx1.0-0.018i$,
$g_{12}\approx0.59-0.005i$, $g_{21}\approx0.06+0.001i$, $L_{\rm
Diff}=L_{\rm NL}\approx1.26$ cm, and the group velocity
\be V_{g}\approx5.6\times10^{-6}\,c. \ee

It is clear that the imaginary parts of the coefficients in Eqs.
(\ref{GNLSE21}) and (\ref{GNLSE22}) are indeed much less than its
real parts. The physical reason of so small imaginary part is due
to the EIT effect induced by the control field
that makes the absorption of the probe field largely suppressed.
In the following discussion, the small imaginary parts of the
coefficients are neglected for analytical analysis, but they are
accounted in numerical simulations.

Note that Eq.~(\ref{GNLSE21}) is valid only for the probe filed
with a large pulse length $\tau_0$ for which group-velocity dispersion effect
of the system can be neglected. To estimate the required order of magnitude of
$\tau_0$, we compare the
characteristic dispersion length (defined by $L_{\rm Disp}={\rm
Re}(\tau_{0}^{2}/|\partial^2 K/\partial \omega^2|_{\omega=0}$) and
the diffraction length $L_{\rm Diff}$ defined above. By setting
$L_{\rm Disp}=L_{\rm Diff}$ we obtain $\tau_{0}=1.48\times10^{-6}$
s. Consequently, if $\tau_0$ is much larger than $1.48\times
10^{-6}$ s, $L_{\rm Disp}$ will be much longer than $L_{\rm Diff}$
and hence the group-velocity dispersion effect of the system can be neglected
safely.

\subsection{Guiding a linear light bullet with one Airy beam}

We first study the possibility of guiding a 3D linear light bullet
with one Airy beam. If $U_0$ is much smaller than $9.0\times10^{6}$
s$^{-1}$, the typical nonlinear length $L_{\rm NL}$ will be much
longer than the typical diffraction length $L_{\rm Diff}$, and
hence $\ell\ll1$. Thus, the SPM term in the Eq.~(\ref{GNLSE21})
can be neglected because $g_{11}\propto\ell\ll1$. However, the
condition $\ell\ll1$ will also suppress the CPM term which
contributes to the trapping potential to the probe field. Without the
potential, if a light bullet is excited, it will be highly
unstable due to the transverse instability \cite{mal,kiv}.
In addition, the potential will also be used to guide the light
bullet. In order to avoid the suppression of the CPM term, we can use a large $v_0$ to
fulfill the condition $\ell v_0^2\sim1$. Taking into account the
above considerations, Eq.~(\ref{GNLSE21}) reduces to a (3+1)D
linear Schr\"{o}dinger equation with a linear potential to the probe field.

Now we turn to the Helmholtz equation (\ref{GNLSE22}). As we know,
it admits different types of centrosymmetric beam solutions such as Gaussian beam,
Bessel beam, and Laguerre-Gaussian beam~\cite{Hall,han3}, etc. However, in this
work we are interested in a particular type of anticentrosymmetric beam solution, i.e.
the Airy beam, with the form $v(s,\xi,\eta)={\rm
Ai}(\xi-s^2/4){\rm
Ai}(\eta-s^2/4)e^{i(\xi/2+\eta/2-s^2/6)s}$~\cite{Berry}. At the
entrance of the medium $v(0,\xi,\eta)={\rm Ai}(\xi){\rm Ai}(\eta)$
which can be experimentally realized by a Gaussian beam passing
through a third-order phase mask. The Airy beam solution has
many striking features. In particular, the intensity profile of its
transverse part remains invariant (i.e. it does
not spread out) when bending along a parabolic trajectory.
However, the Airy beam is  not square integrable (i.e. $\int
Ai^2(x)dx\rightarrow\infty$). One possible way to solve this
problem is to introduce an exponential aperture function, i.e.
$v(0,\xi,\eta)={\rm Ai}(\xi){\rm
Ai}(\eta)e^{a_1\xi+a_2\eta}$~\cite{Siviloglou1,Siviloglou2}. Here
$a_j$ ($j=1,\,2$) are positive parameters introduced to ensure
containment of the infinite Airy tail. Typically, $a_j\ll1$ so
that the resulting profile closely resembles the intended Airy
function. By directly integrating Eq.~(\ref{GNLSE22}) we have
\bea \label{airy} v(s,\xi,\eta)&=&{\rm Ai}(\xi-s^2/4+ia_1\xi){\rm
Ai}(\eta-s^2/4+ia_2\xi)e^{i(\xi/2+\eta/2-s^2/6)s}\nonumber\\
&& \times
e^{a_1\xi-a_1\xi^2/2+ia_1^2\xi/2}e^{a_2\eta-a_2\eta^2/2+ia_2^2\eta/2}.
\eea
It is clear that the center position of the Airy beam (\ref{airy})
moves along the trajectory $\xi=\eta=s^2/4$, and hence tends to
bend itself in transverse directions (i.e. the $x$ and $y$ directions).

Substituting the solution (\ref{airy}) into Eq.~(\ref{GNLSE21})
without the SPM term, we obtain the equation
\bea\label{ODE0}
& &i\left(\f{\pl }{\pl s}+\lambda\f{\pl }{\pl
\tau}\right)u+\f{1}{2}\left(\f{\pl ^{2}}{\pl \xi^{2}}+\f{\pl
^{2}}{\pl \eta^{2}}\right)u+g_{12}
v_0^2|{\rm Ai}(\xi-s^2/4+ia_1\xi) \nonumber\\
& & \times {\rm
Ai}(\eta-s^2/4+ia_2\xi)|^2e^{2a_1\xi-a_1\xi^2}e^{2a_2\eta-a_2\eta^2}u=0.
\eea
We see that the Airy beam provides an ``external potential'' to the
probe field.  Equation (\ref{ODE0}) can be solved by taking~\cite{han1,han2}
\be\label{bullet}
u(\tau,\xi,\eta,s)=\phi(\tau,s)\psi(\tau,\xi,\eta), \ee
with
\be\label{gau} \phi(\tau,s)=\frac{1}{\sqrt[4]{2\pi\rho^2}}
e^{-(s-\tau/\lambda)^2/(4\rho^2)}=\frac{1}{\sqrt[4]{2\pi\rho^2}}e^{-(z-V_g
t)^2/(4\rho^2L_{\rm Diff}^2)}, \ee
where $\rho$ is a free real parameter. When writing Eq.~(\ref{gau})
we have assumed that the probe-field envelope is a
Gaussian pulse propagating in $z$ direction with velocity $V_g$.
In this way, the transverse distribution $\psi(\tau,\xi,\eta)$
satisfies the linear equation
\bea\label{ODE1}
& &i\lambda\f{\pl \psi}{\pl \tau}+\f{1}{2}\left(\f{\pl ^{2}}{\pl
\xi^{2}}+\f{\pl ^{2}}{\pl \eta^{2}}\right)\psi+g_{12}v_0^2|{\rm
Ai}(\xi-s^2/4+ia_1\xi){\rm
Ai}(\eta-s^2/4+ia_2\xi)|^2 \nonumber\\
& & \times e^{2a_1\xi-a_1\xi^2}e^{2a_2\eta-a_2\eta^2}\psi=0,
\eea
and its stationary solutions can be obtained by the transformation
$\psi=\exp{(i\mu \tau)}\tilde{\psi}(\xi,\eta)$, leading to the
linear eigenvalue equation
\be\label{eigen0}
{\cal L}\tilde{\psi}=\lambda\mu\tilde{\psi}, \\
\ee
with the operator
\be {\cal L}=\frac{1}{2}\left(\f{\pl ^{2}}{\pl \xi^{2}}+\f{\pl
^{2}}{\pl \eta^{2}}\right)+g_{12}v_0^2|{\rm Ai}(\xi+ia_1\xi){\rm
Ai}(\eta+ia_2\xi)|^2
e^{2a_1\xi-a_1\xi^2}e^{2a_2\eta-a_2\eta^2},\nonumber
\ee
where $\tilde{\psi}$ is a real function and $\mu$ is the
propagation constant. If $\tilde{\psi}$ is transversely localized, $u$ will be localized in all three
spatial directions and evolve in time. In this way, $u$ will describe a linear light bullet in (3+1)D.

%
\begin{figure}
\centering
\includegraphics[scale=0.5]{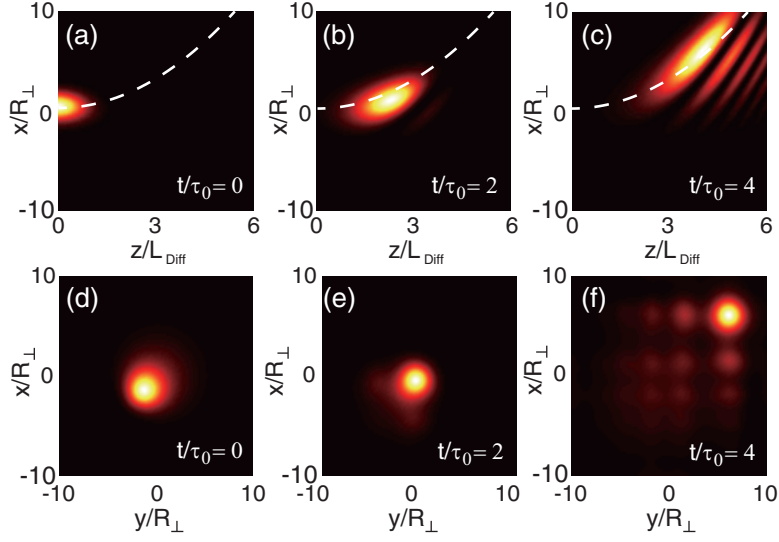}
\caption{\footnotesize (Color online) Guiding a linear light bullet
 with Airy beam ($U_0=1.8\times10^{-6}$ s$^{-1}$). (a)-(c) Intensity patterns of the
linear light bullet in the $x$-$z$ plane at
$\tau/\tau_0=0$, 2, and 4, respectively. Dashed lines denote the trajectory of the main lobe of the Airy
potential. (d)-(f) Intensity patterns
of the linear light bullet in the $x$-$y$ plane at
$\tau/\tau_0=0$, 2, and 4, respectively. A significant diffraction can be observed in (c) and
(f). } \label{fig3}
\end{figure}
%
In Fig.~\ref{fig3} we show the guiding of a typical linear light bullet with the assisted field taken to be an Airy beam. Fig.~\ref{fig3}(a) and Fig.~\ref{fig3}(d) show the intensity pattern of the
linear light bullet by solving Eq.~(\ref{eigen0}). Here we have taken $\tau_{0}=7.5\times10^{-6}$ s so
that $\lambda=1$. To test the stability of the linear light
bullet, we calculate the power of the probe pulse, defined by
$P=2\pi\iiint_{-\infty}^{+\infty}\psi^{2} d\xi d\eta d\tau$, as a
function of the propagation constant $\mu$. For a given $v_0$, $P$
first increases to arrive a maximum, and then decreases. According
to Vakhitov-Kolokolov (VK) criterion \cite{vakhitov1}, the domain
in which the linear light bullet is stable is the one with
$dP/d\mu>0$. Generally, the stability domain is small for small
$v_0$, however, it can be enlarged by increasing $v_0$. This is
because a larger $v_0$ means a stronger trapping to the optical
pulse provided by the potential. In our calculation, the stability domain
is $0<\mu\lesssim0.6$ with $v_0=23$.

The guiding of such linear light bullet is studied by making
simulation of Eq.~(\ref{ODE1}) with the stationary solution in
Fig.~\ref{fig3}(a) and Fig.~\ref{fig3}(d) as the initial condition. The results are
presented in Fig.~\ref{fig4}(b) and Fig.~\ref{fig4}(c) (Fig.~\ref{fig4}(e)
and Fig.~\ref{fig4}(f)\,) at $\tau/\tau_0=2$ and 4,
respectively. We see
that the linear light bullet is indeed guided by the Airy-shaped
assisted field. Specifically, it is trapped in the main lobe of the
Airy beam, propagate ultraslowly in longitudinal direction,
accelerate in transverse directions, and move along a parabolic trajectory.
However, the linear light bullet is unstable because
a significant diffraction occurs during the
propagation, which makes it spread along
the parabolic trajectory and leak energy to the other lobes of the Airy beam (see
Fig.~\ref{fig4}(c) and Fig.~\ref{fig4}(f)\,).

\subsection{Guiding nonlinear light bullets with one Airy beam}

Since the diffraction-induced spreading occurs during the propagation of the
linear light bullet, a natural idea is to use the SPM effect of the
system to balance the  diffraction. To have a significant SPM, one
must increase the amplitude of the probe field. By taking
$U_0=9.0\times10^{6}$ s$^{-1}$ (five times larger than that in the linear case), we have $g_{11}\approx1$ and hence
the SPM term plays an important role in the Eq.~(\ref{GNLSE21}). Substituting the
solution (\ref{airy}) into Eq.~(\ref{GNLSE21}), we obtain the
(3+1)D nonlinear Schr\"{o}dinger (NLS) equation
\bea\label{ODE2}
& &i\left(\f{\pl }{\pl s}+\lambda\f{\pl }{\pl
\tau}\right)u+\f{1}{2}\left(\f{\pl ^{2}}{\pl \xi^{2}}+\f{\pl
^{2}}{\pl \eta^{2}}\right)u+g_{11}|u|^{2}u+g_{12}
v_0^2|{\rm Ai}(\xi-s^2/4+ia_1\xi) \nonumber\\
& & \times {\rm
Ai}(\eta-s^2/4+ia_2\xi)|^2e^{2a_1\xi-a_1\xi^2}e^{2a_2\eta-a_2\eta^2}u=0.
\eea
With (\ref{bullet}) and (\ref{gau}), we have
\bea\label{ODE3}
& &i\lambda\f{\pl \psi}{\pl \tau}+\f{1}{2}\left(\f{\pl ^{2}}{\pl
\xi^{2}}+\f{\pl ^{2}}{\pl
\eta^{2}}\right)\psi+g_{11}|\psi|^{2}\psi+g_{12}
v_0^2|{\rm Ai}(\xi-s^2/4+ia_1\xi){\rm
Ai}(\eta-s^2/4+ia_2\xi)|^2 \nonumber\\
& & \times e^{2a_1\xi-a_1\xi^2}e^{2a_2\eta-a_2\eta^2}\psi=0.
\eea
Similarly, the stationary solutions of Eq.~(\ref{ODE3}) can be obtained by the
transformation $\psi=\exp{(i\mu \tau)}\tilde{\psi}(\xi,\eta)$,
leading to the nonlinear eigenvalue equation
\be\label{eigen1}
{\cal L}\tilde{\psi}+g_{11}\tilde{\psi}^3=\lambda\mu\tilde{\psi}, \\
\ee
where the operator ${\cal L}$ is the same with that defined in Eq.~(\ref{eigen0}).

\begin{figure}
\centering
\includegraphics[scale=0.5]{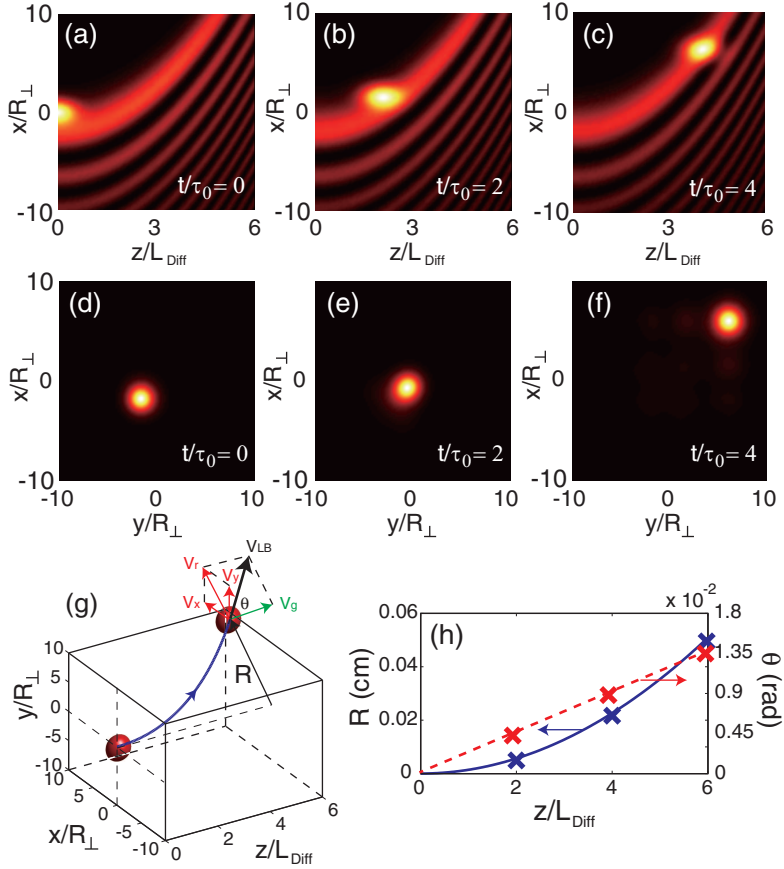}
\caption{\footnotesize (Color online)
Guiding nonlinear light bullet with Airy beam.
(a)-(c) Intensity patterns
of the nonlinear light bullet in the $x$-$z$ plane at
$\tau/\tau_0=0$, 2, and 4, respectively. (d)-(f) Intensity patterns of the
nonlinear light bullet in the $x$-$y$ plane at
$\tau/\tau_0=0$, 2, and 4. (g) Schematic diagram of the propagation of the nonlinear light
bullet in three-dimensional space. $V_x$ ($V_y$) is the velocity in the $x$ ($y$) direction;
$V_r=\sqrt{V_x^2+V_y^2}$ is the radial velocity in the $x$-$y$
plane; $V_g$ is the group velocity in the $z$ direction; $V_{\rm
LB}=\sqrt{V_r^2+V_g^2}$ is the total velocity; $R$ is the transverse displacement after
the nonlinear light bullet passing through the atomic medium; $\theta$ is the angle
between $V_r$ and $V_g$; the (red) solid spheres represents the  nonlinear light bullet.
(h) $R$ and $\theta$ as functions of
$z$. The solid and dashed lines are analytical results, the
``$\times$'' symbols are results by numerical simulations. } \label{fig4}
\end{figure}

Fig.~\ref{fig4}(a) and Fig.~\ref{fig4}(d) show the intensity pattern of a stationary
nonlinear light bullet by solving Eq.~(\ref{eigen1}).  The values of $\tau_0$ and $\rho$ are the same
with those used in the last subsection. For a given $v_0$, the probe-field power $P$
first increases to arrive a maximum, and then decreases. However, the stability domain
of a nonlinear light bullet is larger than that of a linear one. This is
because the focusing nonlinearity favors to the formation of the
nonlinear light bullet, and hence enhances its stability. For example, the
stability domain is $0<\mu\lesssim1.5$ for $v_0=7.3$.

%
%

The guiding of the nonlinear light bullet is studied by making a numerical
simulation of Eq.~(\ref{ODE3}) with the stationary solution given in
Fig.~\ref{fig4}(a) and Fig.~\ref{fig4}(d) as an initial condition. The results in
Fig.~\ref{fig4}(b) and Fig.~\ref{fig4}(c) (Fig.~\ref{fig4}(e) and Fig.~\ref{fig4}(f)\,)
 are for $\tau/\tau_0=2$ and 4, respectively. We see
that the nonlinear light bullet is indeed guided by the
Airy-beam-shaped assisted field. Importantly, different from the linear light
bullet given in the last subsection no evident diffraction is observed
during the propagation of the nonlinear light bullet.
This is because the diffraction is completely balanced by
the SPM effect even the trajectory of the nonlinear light bullet
is bent.

The position of the nonlinear light bullet can be obtained by the
trajectory of the main lobe of the Airy beam, which reads
\be\label{traj} (X, Y, Z)=\left(\f{R_{\perp}V_g^2}{4L_{\rm
Diff}^2}t^2,\f{R_{\perp}V_g^2}{4L_{\rm Diff}^2}t^2,V_gt\right).
\ee
From Eq.~(\ref{traj}) we see that the nonlinear light bullet accelerates
in both $x$ and $y$ directions with the same accelerated velocity
$R_{\perp}V_g^2/(2L_{\rm Diff}^2)$, and propagates in $z$
direction with the constant propagating velocity $V_g$. In a
mechanical point of view, the acceleration of the nonlinear light bullet is
caused by the transverse force produced by the potential
contributed by the assisted field.

For clearance, in Fig.~\ref{fig4}(g) we show the schematic diagram for the propagation
of the nonlinear light bullet in three-dimensional space.
In this figure, $V_x$ and $V_y$ are respectively the velocities
of the nonlinear light bullet in $x$ and $y$ directions,
$V_r=\sqrt{V_x^2+V_y^2}$ is the radial velocity in the transverse plane,
$V_g$ is the velocity in $z$ direction,  $V_{\rm
LB}=\sqrt{V_r^2+V_g^2}$ is the total velocity, $R$ is the transverse displacement after
the light bullet passing through the atomic medium, and $\theta$ is
the angle between $V_r$ and $V_g$ describing the output
direction.

Shown in Fig.~\ref{fig4}(h) are $R$ and $\theta$ as functions of $z$.
The solid and dashed lines are analytical results, while
``$\times$'' symbols are results by making numerical simulation.
We see that the position of the  nonlinear light bullet can be
controlled and manipulated by the Airy beam.
For example, we obtain $R\approx0.05$ cm and
$\theta\approx1.35\times10^{-2}$ rad after the nonlinear light bullet
passing through a medium with the length $6L_{\rm Diff}=7.56$ cm. We
note that the magnitude of the output angle
obtained here is one order larger than
that obtained using a Stern-Gerlach gradient magnetic field
in Ref.~\cite{kar}.

The generation power of the (3+1)D nonlinear light bullet described above
can be estimated by calculating Poynting's vector. The peak power
of the probe field is given by $\bar{P}_{\rm{max}}=2\epsilon_0 c
n_p S_0(\hbar/|\textbf{ p}_{13}|)^2 U_{0}^{2}|u_{\rm max}|^2$,
with $n_p$ and $S_0$ being the reflective index and the
cross-section area of the probe beam, respectively. Taking
$S_{0}=\pi R_{\perp}^{2}\approx 0.5\times 10^{-4}$ cm$^2$ and
using the other parameters given above, we obtain the generation
power of the nonlinear light bullet
\be
\bar{P}_{\rm{max}}\approx 1.8\,\,\mu {\rm W}.
\ee
Consequently, the nonlinear light bullet in the present system
may have not only an ultraslow propagating velocity but also a very low generation power.
This is fundamentally different from the other generation schemes where
the light bullets have the propagating velocity of the same order of $c$
and their generation power up to megawatt is needed
\cite{tra,min}.

\subsection{Guiding nonlinear light bullets with two sequential Airy beams}

In this subsection we show that if the assisted field is taken to be two sequential Airy beams,
the nonlinear light bullet can easily bypass an obstacle. To this end,
we assume that the assisted field takes the form
\bea \label{airyairy} v(\tau,s,\xi,\eta)&=&\{f_1(\tau){\rm
Ai}(\xi-s^2/4+ia_1\xi){\rm
Ai}(\eta-s^2/4+ia_2\xi)e^{i(\xi/2+\eta/2-s^2/6)s}\nonumber\\
&& +f_2(\tau){\rm Ai}[\xi-(s-s_0)^2/4+ia_1\xi]{\rm
Ai}[\eta-(s-s_0)^2/4+ia_2\xi]\nonumber\\
&& \times e^{i[\xi/2+\eta/2-(s-s_0)^2/6](s-s_0)}\}
e^{a_1\xi-a_1\xi^2/2+ia_1^2\xi/2}e^{a_2\eta-a_2\eta^2/2+ia_2^2\eta/2},
\eea
where
$f_1(\tau)=\frac{1}{2}\left\{1-\tanh[2(\tau-\tau_1)]\right\}$ and
$f_2(\tau)=\frac{1}{2}\left\{1+\tanh[2(\tau-\tau_1)]\right\}$, with $s_0$ being the length of the medium.
Clearly, the solution (\ref{besselairy}) obeys the Helmholtz equation
(\ref{GNLSE22}) because it is a combination of two sequential Airy beams,
propagating respectively along $z$ and $-z$ directions in different time.

%
\begin{figure}
\centering
\includegraphics[scale=0.5]{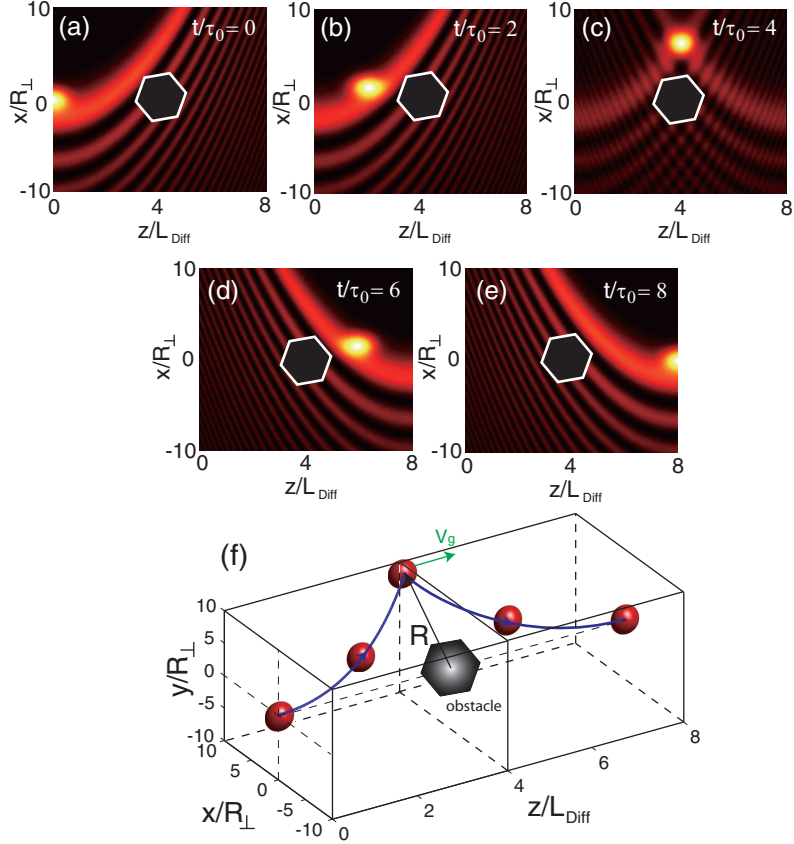}
\caption{\footnotesize (Color online) Nonlinear light bullet
bypasses an obstacle.  (a)-(e) Intensity patterns of the nonlinear
light bullet in the $x$-$z$ plane for $\tau/\tau_0=0$, 2, 4, 6,
and 8, respectively. The black Polygone  represents an obstacle.
(f) The propagation of the nonlinear light bullet (represented by
the (red) solid spheres) in the 3D space. The ``$\Lambda$'' shape
trajectory of the nonlinear light bullet enables it to bypass the
obstacle (represented by the black Polygone).} \label{fig5}
\end{figure}
%

In Fig.~\ref{fig5}(a)-Fig.~\ref{fig5}(e) we show the intensity
patterns of the nonlinear light bullet at $t/\tau_0=0$, 2, 4, 6,
and 8, respectively, for $\tau_1=4$ and $s_0=8$. In the first time
interval, i.e. $\tau\in(0,4)$, the nonlinear light bullet with
initial position $(x,y,z)=(0,0,0)$ is trapped in the forward Airy
beam (i.e. the beam with $f_1(\tau)\approx 1$, $f_2(\tau)\approx
0$) and moves along the main lobe of the beam to the position
$(x,y,z)=(4R_{\perp},4R_{\perp}, 4L_{\rm Diff})$ at $\tau=4$, as
shown in Fig.~\ref{fig5}(c). In the second time interval, i.e.
$\tau\in(4,8)$, the forward Airy beam is switched off and the
backward Airy beam (i.e. the beam with $f_1(\tau)\approx 0$,
$f_2(\tau)\approx 1$) is switched on. In this time interval, the
nonlinear light bullet is trapped in the backward Airy beam and
moves along the main lobe of the beam to the position
$(x,y,z)=(4R_{\perp},4R_{\perp}, 8L_{\rm Diff})$ at $\tau=8$, as
shown in Fig.~\ref{fig5}(e). Interestingly, we see that the
nonlinear light bullet travels along a ``$\Lambda$'' shape
trajectory. Consequently, if there is an obstacle which is put in
the position below the ``$\Lambda$'' shape trajectory, the
nonlinear light bullet can bypass the obstacle, as shown in
Fig.~\ref{fig5}(f) (in all panels, the black Polygone represents
the obstacle).

\section{Generation of nonlinear helical light bullets}

The Airy beam can also be used to
generate an ultraslow helical weak-light bullet proposed in Ref.~\cite{han2}.
To this end, we assume that the assisted field takes the form
\bea \label{besselairy}
v(\tau,s,\xi,\eta)&=&f_1(\tau){\rm
Ai}(\xi-s^2/4+ia_1\xi){\rm
Ai}(\eta-s^2/4+ia_2\xi)e^{i(\xi/2+\eta/2-s^2/6)s}\nonumber\\
&& \times
e^{a_1\xi-a_1\xi^2/2+ia_1^2\xi/2}e^{a_2\eta-a_2\eta^2/2+ia_2^2\eta/2}+f_2(\tau)J_1(\sqrt{2b}r),
\eea
with
$f_1(\tau)$ and $f_2(\tau)$ being the same with those defined in Eq.~(\ref{airyairy}),
$J_1$ being the first-order Bessel function, $b$ being
a real constant characterizing the radius of the Bessel
function, and $r=\sqrt{\xi^2+\eta^2}$. It is clear that the
solution (\ref{besselairy}) also obeys the Helmholtz equation
(\ref{GNLSE22}) because it is a combination of Bessel and Airy
beams which are both solutions of the Helmholtz equation.

%
\begin{figure}
\centering
\includegraphics[scale=0.5]{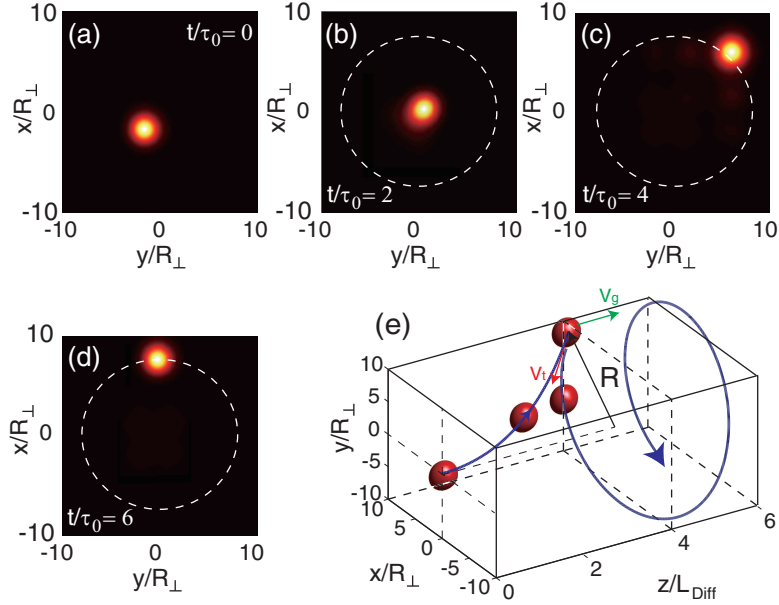}
\caption{\footnotesize (Color online) Generating a nonlinear helical light
bullet with Airy beam. (a)-(d) Intensity patterns of the nonlinear light bullet
in the $x$-$y$ plane at $t/\tau_0=0$, 2, 4, and 6, respectively, for  $\tau_1=4$ and $b=0.05$.
(e) The propagation of the helical light bullet in 3D space. The solid line with arrow denotes
the motion trajectory of the nonlinear light bullet.
The radius of the ring $R\approx5.66R_{\perp}$.} \label{fig6}
\end{figure}
%

In Fig.~\ref{fig5}(a)-Fig.~\ref{fig5}(d) we show the intensity patterns of the light bullet at
$t/\tau_0=0$, 2, 4, and 6, respectively,  for $\tau_1=4$ and $b=0.05$.
In the first time interval, $\tau\in(0,4)$, the nonlinear light bullet
is trapped in the Airy beam and moves to the position
$(x,y)=(4R_{\perp},4R_{\perp})$ at the end of the first interval,
as shown in Fig.~\ref{fig6}(c). After the first time interval, we switch off the Airy beam
and switch on the first-order Bessel beam, the nonlinear light bullet is
then trapped in the first ring of the first-order Bessel beam and
moves along the ring if the trapping potential contributed by the
ring is narrow and deep enough. This is possible because in each
ring of the Bessel beam the potential energy is degenerate and
reaches its minimum, therefore a light bullet will move along the
ring if an initial transverse velocity $V_t$ tangent to the ring
is given. Notice that after switching off the Airy beam the
velocity of the light bullet in the transverse plane is the radial
velocity $V_r$ which is orthogonal to the ring, and hence it can
not trigger on the rotary motion when the first-order Bessel beam
is switched on. However, a tangent velocity $V_t$ can be produced
by various methods such as using a gradient magnetic field \cite{han2} or a
shift of the Bessel lattice~\cite{he}. Then, the
nonlinear light bullet rotates around the circle, as shown in Fig.~\ref{fig6}(d).

Since now the nonlinear light bullet has two orthogonal velocities, the tangent velocity $V_t$ and the
group velocity $V_g$, they can actually make a helical
motion in the 3D space, as shown in Fig.~\ref{fig6}(e), where
the solid line with arrow denotes
the motion trajectory of the nonlinear light bullet. Because
both velocities are much smaller than $c$ and the generation power
of the nonlinear light bullet is very weak,
such light bullet is named as the ultraslow helical weak-light bullet.

In general, it is possible to move a nonlinear light bullet from the
center of the transverse plane to any ring of Bessel lattices.
By using such assisted field with sequential Airy and Bessel beams,
one can manipulate and control the output position of a nonlinear light bullet in a very
efficient way.

\section{Summary}

In this article, we have studied the possibility of guiding stable
ultraslow weak-light bullets by using Airy beams in a cold,
lifetime-broadened four-level atomic system via EIT. We have shown
that under the EIT condition the light bullet with ultraslow
propagating velocity ($\sim 10^{-5}\,c$) and extremely low
generation power ($\sim 1\,\mu W$) formed by the balance between
diffraction and nonlinearity in the probe field can be not only
stabilized but also guided by the assisted field. In particular,
when the assisted field is taken to be an Airy beam the light
bullet can be trapped into the main lobe of the Airy beam,
propagate ultraslowly in longitudinal direction, accelerate in
transverse directions, and hence move along a parabolic
trajectory. We have demonstrated that the light bullet can bypass
an obstacle by using two sequential Airy beams. A technique of
generating ultraslow helical weak-light bullets in the present
system has also been proposed. The results obtained in this work
are useful for guiding new experimental findings and have
potential applications in optical information processing and
transmission. For instance, the guided light bullets suggested
here can be used to design all-optical switching and logic gates.
In addition, they can also be employed to design new type of
all-optical routers for transmitting optical information.


\acknowledgments This work was supported by the NSF-China under
Grant Numbers 11174080 and 11105052.

\appendix

\section{Equations of motion for $\sigma_{ij}$}

Equations of motion for $\sigma_{ij}$ are given by
\begin{subequations}
\label{den}
\bea
& & i\f{\pl}{\pl
t}\sigma_{11}-i\Gamma_{31}\sigma_{33}+\Omega_{p}^{\ast}
\sigma_{31}-\Omega_{p}\sigma_{31}^{\ast}=0,\label{den1}\\
& & i\f{\pl}{\pl
t}\sigma_{22}-i\Gamma_{32}\sigma_{33}-i\Gamma_{42}\sigma_{44}+\Omega_{c}^{\ast}
\sigma_{32}-\Omega_{c}\sigma_{32}^{\ast}+\Omega_{a}^{\ast}
\sigma_{42}-\Omega_{a}\sigma_{42}^{\ast}=0,\label{den2}\\
& & i\left( \f{\pl}{\pl t}+\Gamma_{3}\right) \sigma_{33}
    -\Omega_{p}^{\ast}
\sigma_{31}+\Omega_{p}\sigma_{31}^{\ast}-\Omega_{c}^{\ast}
\sigma_{32}+\Omega_{c}\sigma_{32}^{\ast}=0,\label{den3}\\
& & i\left( \f{\pl}{\pl t}+\Gamma_{4}\right) \sigma_{44}
    -\Omega_{a}^{\ast}
\sigma_{42}+\Omega_{a}\sigma_{42}^{\ast}=0,\label{den4}\\
& & \left(i\f{\pl}{\pl
t}+d_{21}\right)\sigma_{21}+\Omega_{c}^{\ast}\sigma_{31}
+\Omega_{a}^{\ast}\sigma_{41}-\Omega_{p}\sigma_{32}^{\ast}
=0,\label{den21}\\
& & \left(i\f{\pl}{\pl
t}+d_{31}\right)\sigma_{31}+\Omega_{p}(\sigma_{11}-\sigma_{33})
+\Omega_{c}\sigma_{21}
=0,\label{den31}\\
& & \left(i\f{\pl}{\pl
t}+d_{41}\right)\sigma_{41}+\Omega_{a}\sigma_{21}
-\Omega_{p}\sigma_{43}=0,\label{den41}\\
& & \left(i\f{\pl}{\pl
t}+d_{32}\right)\sigma_{32}+\Omega_{c}(\sigma_{22}-\sigma_{33})
+\Omega_{p}\sigma_{21}^{\ast}-\Omega_{a}\sigma_{43}^{\ast}
=0,\label{den32}\\
& & \left(i\f{\pl}{\pl
t}+d_{42}\right)\sigma_{42}+\Omega_{a}(\sigma_{22}-\sigma_{44})
-\Omega_{c}\sigma_{43}=0,\label{den42}\\
& & \left(i\f{\pl}{\pl
t}+d_{43}\right)\sigma_{43}+\Omega_{a}\sigma_{32}^{\ast}
-\Omega_{p}^{\ast}\sigma_{41}-\Omega_{c}^{\ast}\sigma_{42}
=0,\label{den43}
\eea
\end{subequations}
where $\Gamma_{ij}$ is the rate at which population decays from the state
$|i\rangle$ to the state $|j\rangle$,
$d_{ij}=\Delta_i-\Delta_j+i\gamma_{ij}$ with
$\gamma_{ij}\equiv (\Gamma_{i}+\Gamma_{j})/2+\gamma^{{\rm dph}}_{ij}$. Here
$\Gamma_i=\sum_{E_j<E_i}\Gamma_{ij}$ and $\gamma_{ij}^{\text{col}}$
denotes the dipole dephasing rate caused by atomic collisions.

\section{Explicit expressions of $\alpha_{jl}$}

The explicit expressions of $\alpha_{jl}$ read

\bes \label{alp}
\bea
& & \label{al11} \alpha_{11}=\f{\kappa_{13}}{D}\left\{\Omega_{c}a_{32}^{\ast (2)}
-(\omega+d_{21})\left[\f{4}{\Gamma_{31}}{\rm Im}\left(\f{d_{21}}{D}\right)
+a_{22}^{(2)}\right]\right\},\\
& & \label{al12} \alpha_{12}=-\f{\kappa_{13}|\Omega_{c}|^{2}}{(\omega+d_{41})D^{2}},\\
& & \alpha_{21}=\f{\kappa_{24}}{|\Omega_{c}|^{2}-d_{42}d_{43}}\left[d_{43}a_{22}^{(2)}
+\Omega_{c}a_{32}^{\ast(2)}-\f{|\Omega_{c}|^{2}}{(\omega+d_{41})D}\right],
\eea
\ees
with
\bes \label{a}
\bea
& & a_{22}^{(2)}=\left[\f{2}{\Gamma_{31}}{\rm
Im}\left(\f{d_{21}}{D}\right)- \f{{\rm
Im}\left(\f{1}{d_{32}^{\ast}D}\right)}{{\rm
Im}\left(\f{1}{d_{32}}\right)}-\f{\Gamma_{32}}{\Gamma_{31}|\Omega_{c}|^{2}}
\f{{\rm Im}\left(\f{d_{21}}{D}\right)}{{\rm
Im}\left(\f{1}{d_{32}}\right)}\right],\\
& & a_{33}^{(2)}=\f{2{\rm
Im}\left(\f{d_{21}}{D}\right)}{\Gamma_{31}}\\
& &
a_{32}^{(2)}=\f{1}{d_{32}}\left[\f{\Omega_{c}}{D^{\ast}}
+\Omega_{c}(a_{33}^{(2)}-a_{22}^{(2)})\right].
\eea
\ees
%




\begin{references}

\bibitem{sil}Y. Silberberg, Opt. Lett. {\bf 22}, 1282 (1990).

\bibitem{ber}L. Berge, Phys. Rep. {\bf 303}, 260 (1998).

\bibitem{kiv0}Y. S. Kivshar and D. E. Pelinovsky, Phys. Rep. {\bf 331},
             117 (1998).

\bibitem{mal}B. A. Malomed, D. Mihalache, F. Wise, and L. Torner,
J. Phys. B: Quantum Semiclass. Opt. {\bf 7}, R53 (2005), and references therein.

\bibitem{kiv}Y. S. Kivshar and G. P. Agrawal, {\it Optical
Solitons:  From Fibers to Photonic Crystals} (Academic Press, London, 2006),
and references therein.

\bibitem{liu}X. Liu, L. J. Qian and F. W. Wise,
Phys. Rev. Lett. {\bf 82}, 4631 (1999).

\bibitem{bla}M. Blaauboer, B. A. Malomed, and G. Kurizki,
Phys. Rev. Lett. {\bf 84}, 1906 (2000).

\bibitem{tow}I. N. Towers, B. A. Malomed, and F. W. Wise,
Phys. Rev. Lett. {\bf 90}, 123902 (2003).

\bibitem{tra}P. D. Trapani, G. Valiulis, A. Piskarskas, O.
Jedrkiewicz, J. Trull, C. Conti, and S. Trillo,  Phys. Rev. Lett.
{\bf 91}, 093904 (2003).

\bibitem{mih}D. Mihalache, D. Mazilu, F. Ledererm B. A.
Malomed, Y. V. Kartashov, L.-C. Crasovan, and L. Torner, Phys.
Rev. Lett. {\bf 95}, 023902 (2005).

\bibitem{mat}M. Matuszewski, E. Infeld, B. A. Malomed, and M.
Trippenbach,  Phys. Rev. Lett. {\bf 95}, 050403 (2005).

\bibitem{ber1}L. Berg\'{e} and S. Skupin,
Phys. Rev. Lett. {\bf 100}, 113902 (2008).

\bibitem{bel}M. Beli\'{c}, N. Petrovi\'{c}, W. P. Zhong,
R. H. Xie, and G. Chen, Phys. Rev. Lett. {\bf 101}, 123904 (2008).

\bibitem{burg}I. B. Burgess, M. Peccianti, G. Assanto, and R.
Morandotti, Phys. Rev. Lett. {\bf 102}, 203903 (2009).

\bibitem{chen}S. H. Chen and J. M. Dudley,
Phys. Rev. Lett. {\bf 102}, 233903 (2009).


\bibitem{abd}D. Abdollahpour, S. Suntsov, D. G. Papazoglou, and
S. Tzortzakis, Phys. Rev. Lett. {\bf 105}, 253901 (2010).

\bibitem{min}S. Minardi, F.
Eilenberger, Y. V. Kartashov, A. Szameit, U. R\"{o}pke, J.
Kobelke, K. Schuster, H. Bartelt, S. Nolte, L. Torner, F. Lederer,
A. T\"{u}nnermann, and T. Pertsch, Phys. Rev. Lett. {\bf 105},
263901 (2010).

\bibitem{mat1}A. M. Mateo, V. Delgado, and B. A. Malomed,
Phys. Rev. A {\bf 82}, 053606 (2010).

\bibitem{kat1}Y. V. Kartashov, B. A. Malomed, and L. Torner,
Rev. Mod. Phys. {\bf 83}, 247 (2011).


\bibitem{mihalache1}D. Mihalache, D. Mazilu, F. Lederer, and Y. S.
Kivshar, Opt. Lett. {\bf 32}, 3173 (2007).

\bibitem{mihalache2}D. Mihalache, D. Mazilu, F. Lederer, and Y. S.
Kivshar, Phys. Rev. A {\bf 79}, 013811 (2009).

\bibitem{mihalache3}D. Mihalache, J. Opt. Adv. Mat.  {\bf 12}, 12 (2010).



\bibitem{har} S. E. Harris, Phys. Today {\bf 50}(7), 36 (1997).

\bibitem{fle} M. Fleischhauer, A. Imamoglu, and J. P. Marangos,
Rev. Mod. Phys. {\bf 77}, 633 (2005), and references therein.


\bibitem{wu}Y. Wu and L. Deng, Phys. Rev. Lett. {\bf 93},
143904 (2004).

\bibitem{hua}G. Huang, L. Deng and M. G. Payne,
Phys. Rev. E. {\bf 72}, 016617 (2005).

\bibitem{hang}C. Hang and G. Huang,
Phys. Rev. A {\bf 77}, 033830 (2008).

\bibitem{yang}W.-X. Yang, A.-X. Chen, L.-G. Si, K. Jiang, X. Yang, and R.-K.
Lee,  Phys. Rev. A {\bf 81}, 023814 (2010).



\bibitem{hong}T. Hong, Phys. Rev. Lett. {\bf 90}, 183901 (2003).

\bibitem{mic}H. Michinel and M. J. Paz-Alonso,
Phys. Rev. Lett. {\bf 96}, 023903 (2006).

\bibitem{hang1}C. Hang, G. Huang, and L. Deng,
Phys. Rev. E {\bf 73}, 046601 (2006).

\bibitem{hang2}C. Hang, V. V. Konotop, and G. Huang,
 Phys. Rev. A {\bf 79}, 033826 (2009).



\bibitem{LWH}H. Li, Y. Wu, and G. Huang, Phys. Rev. A {\bf 84}, 033816 (2009).


\bibitem{han1}C. Hang and G. Huang, Phys. Rev. A {\bf 86}, 043809 (2012).

\bibitem{han2}C. Hang and G. Huang, Phys. Rev. A {\bf 87}, 053809 (2013).



\bibitem{Berry} M. V. Berry and N. L. Balazs, Am. J. Phys. {\bf 47}, 264 (1979).

\bibitem{Bandres}M. A. Bandres, I. Kaminer, M. S. Mills, B. M. Rodriguez-Lara,  E. Greenfield,
M. Segev, and D. N. Christodoulides,   Opt. \& Photon. News {\bf
24}, 30 (2013).

\bibitem{bau}J. Baumgartl, M. Mazilu, and K. Dholakia, Nature Photonics {\bf 2}, 675 (2008).

\bibitem{zhang}P. Zhang, J. Prakash, Z. Zhang, M. S. Mills, N. K. Efremidis, D. N. Christodoulides,
and Z. Chen, Opt. Lett. {\bf 36}, 2883 (2011).

\bibitem{Polynkin}P. Polynkin, M. Kolesik, J. V. Moloney, G. A. Siviloglou, and D. N. Christodoulides,
Science {\bf 324}, 229 (2009).

\bibitem{zhu}L. Li, T. Li, S. M. Wang, C. Zhang, and S. N. Zhu, Phys. Rev. Lett. {\bf 107}, 126804 (2011).


\bibitem{note0}Here the first `3' refers to spatial coordinates
and `1' refers one time coordinate.

\bibitem{steck}D. A. Steck, ``Rubidium 87 D Line Data'',
http://steck.us/alkalidata/.


\bibitem{note1}
The frequency and wavevector of the probe field is given by
$\omega_p+\omega$ and $k_p+K_p(\omega)$. Thus $\omega=0$ corresponds to
the center frequency of the probe field.

\bibitem{lil}L. Li and G. Huang, Phys. Rev. A {\bf 82}, 023809 (2010).


\bibitem{Hall}D. G. Hall, Opt. Lett. {\bf 21}, 9 (1996).

\bibitem{han3}C. Hang and V. V. Konotop, Phys. Rev. A {\bf 83}, 053845 (2012).


\bibitem{Siviloglou1} G. A. Siviloglou and D. N. Christodoulides, Opt. Lett. {\bf 32},
979 (2007).

\bibitem{Siviloglou2} G. A. Siviloglou, J. Broky, A. Dogariu, and  D. N. Christodoulides,
Phys. Rev. Lett. {\bf 99}, 213901 (2007).

\bibitem{vakhitov1}M.
G. Vakhitov and A. A. Kolokolov, Sov. J. Radiophys. Quantum
Electron. {\bf 16}, 783 (1973).

\bibitem{kar} L. Karpa and M. Weitz, Nat. Phys. {\bf 2}, 332 (2006).

\bibitem{he}Y. J. He, Boris A. Malomed, and H. Z. Wang, Phys. Rev. A {\bf 76}, 053601 (2007).

\end{references}
\end{document}